\documentclass[aps,prc,superscriptaddress,showpacs,amsmath,amssym,
nofootinbib,
twocolumn]{revtex4-1}

\usepackage{array}
\usepackage{graphicx}
\usepackage{epstopdf}
\usepackage{color}
\usepackage{bm}

\newcommand{\beq}{\begin{equation}}
\newcommand{\eeq}{\end{equation}}

\begin{document}

\title{Diffusion and Coulomb separation of ions in dense matter}
\author{M.\ V.\ Beznogov}
\affiliation{St.~Petersburg Academic University, 8/3 Khlopina Street,
St.~Petersburg 194021, Russia}
\author{D.\ G.\ Yakovlev}
\affiliation{Ioffe Physical Technical Institute, 26 Politekhnicheskaya,
St.~Petersburg 194021, Russia}

\date{October 14, 2013}

\begin{abstract}
We analyze diffusion equations in strongly coupled Coulomb mixtures
of ions in dense stellar matter. Strong coupling of ions in the
presence of gravitational forces and electric fields (induced by
plasma polarization in the presence of gravity) produces a specific
diffusion current which can separate ions with the same $A/Z$
(mass to charge number) ratios but different $Z$. This Coulomb
separation of ions can be important for the evolution of white dwarfs and
neutron stars.
\end{abstract}

\maketitle

\section{Introduction}
\label{s:introduct}

In this Letter we consider diffusion in a multicomponent plasma of ions
in dense stellar matter. This diffusion can greatly affect the
composition of matter particularly in cores of white dwarfs
and envelopes of neutron stars. It produces redistribution of ions
(heavier ions move to the star's center) and extra energy release
that reheats the star and affects its thermal evolution. For example,
we can mention gravitational settling of $^{22}$Ne in carbon-oxygen
($^{12}$C--$^{16}$O) cores of white dwarfs (see, e.g., Refs.\
\cite{isernetal91,bh01,db02,althausetal10,garciaetal10})which
is thought to reheat old white dwarfs and helps explain
observational data. Diffusion of ions affects also chemical
evolution and nuclear burning in envelopes of neutron stars (e.g.,
Refs.\ \cite{cb03,cb04,cb10}). Transport properties are important
also in dusty plasmas with their numerous applications (e.g., Ref.\
\cite{dusty}).

Diffusion equations are well studied in physical kinetics for the
case when the ions constitute almost ideal plasma \cite{cc52,hirsh}.
However, the ion plasma in white dwarfs and the neutron
stars is typically strongly coupled by Coulomb
forces. The diffusion coefficients of ions in strongly coupled
Coulomb plasma have been extensively studied in the literature,
mainly by molecular dynamics simulations (e.g., Refs.\
\cite{hansenetal75,horowitz10,horowitz11}).
Here we address a delicate problem of diffusion
currents in strongly coupled Coulomb plasma of ions.

\section{Diffusion currents}
\label{s:Diffusion currents}

Consider a plasma which consists of electrons and a mixture of ion
species $j=1,2,\ldots$, with atomic numbers $A_j$ and charge numbers
$Z_j$. Let $n_j$ be the number density of ions $j$. The electron
number density is $n_e = \sum_j Z_j n_j$ (due to charge neutrality).

It is convenient to introduce (e.g., Ref.\ \cite{hpy07}) the Coulomb
coupling parameter $\Gamma_j$ for ions $j$,
\begin{equation}
 \Gamma_j={ Z_j^2 e^2 \over a_j k_{B} T}
   ={Z_j^{5/3}e^2 \over a_e k_{B} T} ,
\label{Gammaj}
\end{equation}
where $T$ is the temperature, $k_{B}$ is the Boltzmann constant,
$a_e=(4\pi n_e/3)^{-1/3}$ is the electron-sphere radius, and $a_j=a_e
Z_j^{1/3}$ is the ion-sphere radius (for a sphere around a given ion,
where the electron charge compensates the ion charge). Therefore,
$\Gamma_j$ is the ratio of a typical electrostatic energy of the ion
to the thermal energy. If $\Gamma_j \ll 1$ then the ions constitute
an almost ideal Boltzmann gas, while for $\Gamma_j \gtrsim 1$ they
are strongly coupled by Coulomb forces (constitute either Coulomb
liquid or solid). One component ion plasma solidifies at
$\Gamma \approx 175$. We restrict ourselves to the gaseous or liquid
ion plasma.

The diffusion currents in almost ideal plasma are well
defined  \cite{cc52,hirsh} but the case of nonideal plasmas 
requires special attention. We
introduce these currents in the spirit of Landau and
Lifshitz \cite{ll87}. Consider the matter which is slightly off
thermodynamic equilibrium because of the presence of forces
$\bm{F}_\alpha$ [which act on all particles $\alpha$ -- electrons
($\alpha=e$) and ions ($\alpha=j$)] and
number density gradients $\bm{\nabla}n_\alpha$. For simplicity consider
isothermal matter (no temperature gradients
$\bm{\nabla}T=0$, and hence no deviations from thermal equilibrium).
The forces $\bm{F}_\alpha$ and gradients $\bm{\nabla}n_\alpha$ induce (weak)
gradients of chemical potentials $\bm{\nabla}\mu_\alpha$ of particles $\alpha$.
Let us
introduce %
\begin{equation}
  \widetilde{\bm{F}}_\alpha=\bm{F}_\alpha-\bm{\nabla}\mu_\alpha=
  e_\alpha \bm{E} + m_\alpha \bm{g}-\bm{\nabla}\mu_\alpha,
\end{equation}
where we set $\bm{F}_\alpha=e_\alpha\bm{E}+m_\alpha\bm{g}$
($e_\alpha$ and $m_\alpha$ being electric charge and mass of
particles $\alpha$, respectively). The force $\bm{F}_\alpha$ is
produced by gravitational acceleration $\bm{g}$ (that can be treated
as a constant in the local approximation) and the electric field
$\bm{E}$ due to weak plasma polarization in the gravitational field.

Note that
\begin{equation}
   \sum_\alpha n_\alpha \widetilde{\bm{F}}_\alpha=\rho \bm{g}-\bm{\nabla}P,
\end{equation}
where $\rho=\sum_\alpha m_\alpha n_\alpha$ is the mass density of
the matter, and $\bm{\nabla}P=\sum_\alpha n_\alpha \bm{\nabla}
\mu_\alpha$ (as prescribed by thermodynamics \cite{llstat}), $P$
being the (total) pressure. The electric field drops off the sum
because of electric neutrality but it is most important for driving
different particle species (e.g., Refs.\ \cite{cb04,cb10}).

If particles $\alpha$ are in mechanical equilibrium, then
$\widetilde{\bm{F}}_\alpha$=0. This condition is exactly the same as
the condition of chemical equilibrium used in Ref.\ \cite{cb10}. If
the plasma is in hydrostatic equilibrium as a whole, then
$\sum_\alpha n_\alpha \widetilde{\bm{F}}_\alpha=0$ and $\rho
\bm{g}=\bm{\nabla}P$. Hydrostatic equilibration ($\rho
\bm{g}=\bm{\nabla}P$) in neutron stars and white dwarfs is
established over milliseconds---minutes \cite{st83} but diffusive
motion of ions can last over gigayears (Gyrs) (see, e.g., Ref. \cite{db02}). This
diffusion is studied by standard methods of physical kinetics
\cite{cc52,hirsh} assuming $\rho \bm{g}=\bm{\nabla}P$.

In the diffusion problem, a deviation of particles $\alpha$ from
mechanical equilibrium in matter can be conveniently measured by the
vector
\begin{equation}
     {\bm{d}}_\alpha=
     \frac{\rho_\alpha}{\rho}\,\sum_\beta n_\beta \widetilde{\bm{F}}_\beta - n_\alpha
     \widetilde{\bm{F}}_\alpha ,
\label{e:d}
\end{equation}
where $\rho_\alpha=m_\alpha n_\alpha$ is the partial mass density of particles $\alpha$.
Clearly, $\sum_\alpha {\bm{d}}_\alpha=0$.

Let $\bm{J}_\alpha=\rho_\alpha \bm{V}_\alpha$ be the diffusive flux
of particles $\alpha$ ($\bm{V}_\alpha$ being the diffusion velocity
of particles $\alpha$ \cite{cc52,hirsh}). Phenomenological transport
equations can be written as
\begin{equation}
      \bm{J}_\alpha=\Phi \sum_{\beta \neq \alpha}m_\alpha m_\beta
      {D}_{\alpha \beta}{\bm{d}}_\beta,
 \label{e:Ja}
\end{equation}
where ${D}_{\alpha \beta}$ [cm$^2$ s$^{-1}$] can be called a generalized
diffusion coefficient of particles $\alpha$ relative to $\beta$, and $\Phi$ is
a normalization function to be chosen later. The
diffusion coefficients should respect the relation
\begin{equation}
    \sum_\alpha \bm{J}_\alpha=0.
\label{e:sumJa}
\end{equation}

In a rarefied, almost ideal plasma, we have
$\widetilde{\bm{F}}_\alpha=\bm{F}_\alpha-n_\alpha^{-1}
\bm{\nabla}P_\alpha$, where $P_\alpha$ is the partial pressure of
particles $\alpha$. Then Eq.\ (\ref{e:Ja}) reduces to the standard
definition of diffusion coefficients in rarefied gases
\cite{cc52,hirsh}. For strongly interacting particles, partial
pressures $P_\alpha$ are ambiguous, while the definition
(\ref{e:Ja}), based on chemical potentials $\mu_\alpha$, is not.

While the ions are heavy and slow, the electrons are light and
mobile. If we are interested in transport properties of ions, we can
describe the electrons by the approximation similar to the
Born-Oppenheimer approximation in the theory of molecules
\cite{schiff}. Specifically, we assume that the electron gas is
always in the state of mechanical (quasi)equilibrium adjusting
itself almost instantly to the motion of multicomponent ion plasma.
Since the electrons are light, we can set $m_e \to 0$. Then from
Eq.\ (\ref{e:d}) we have ${\bm d}_e=-n_e\,\widetilde{\bm{F}}_e$.
Therefore, the electron quasiequilibrium implies
\begin{equation}
  {\bm d}_e=0,\quad \widetilde{\bm{F}}_e=-e
  \bm{E}-\bm{\nabla}\mu_e=0.
\label{e:electron-equil}
\end{equation}
It allows us to factorize electrons out in the problem of ion
transport (diffusive fluxes of ions are mostly determined by a
nonequilibrium state of the ion subsystem \cite{paquette}). In this
case Eqs.\ (\ref{e:Ja}) and (\ref{e:sumJa}) retain their form but
indices $\alpha$ and $\beta$ label only ion species
($j=1,2,\ldots$). Note that Eq.\ (\ref{e:Ja}) is strictly valid for
nonrelativistic particles, whereas the electrons in dense matter
can be relativistic. However, the factorization works well even for
relativistic electrons as long as they can be treated as massless.

In the presence of two ion species ($j$=1, 2) we have
$\bm{J}_1=-\bm{J}_2$, ${\bm{d}}_1=-{\bm{d}}_2$, and
${D}_{12}={D}_{21}\equiv D$. Then kinetic phenomena can be
characterized by one diffusion coefficient $D$,
\begin{equation}
    \bm{J}_2=-\bm{J}_1=\frac{nm_1 m_2}{\rho k_BT}\,
      {D}{\bm{d}}_1.
\label{e:J1}
\end{equation}
Here we have chosen $\Phi=n/(\rho k_B T)$ ($n=n_1+n_2$ being the
total number density of the ions). Then $D$ corresponds
to the standard definition of the diffusion coefficient
\cite{cc52,hirsh} for two-component plasma of ions (as follows from
the equations presented below). Let us simplify Eq.\ (\ref{e:J1}).

From Eq.\ (\ref{e:d}) we have
\begin{equation}
    {\bm{d}}_j=
     -\frac{\rho_j}{\rho}\bm{\nabla}P-
     n_j e Z_j\bm{E}+n_j \bm{\nabla}\mu_j,
\label{e:d1}
\end{equation}
with $j=1$ or 2. Because $\bm{d}_1+\bm{d}_2=0$, we obtain the
expression for $\bm{E}$:
\begin{equation}
    en_e\bm{E}=-\bm{\nabla}P+
    n_1 \bm{\nabla}\mu_1+n_2 \bm{\nabla}\mu_2.
\end{equation}

Substituting it into (\ref{e:d1}) and setting $m_j=A_jm_u$
($m_u$ being the atomic mass unit), we have
\begin{eqnarray}
  {\bm{d}}_1=\frac{n_1 n_2}{n_e} \,\left[
  m_u(Z_1A_2-Z_2A_1)\,\frac{\bm{\nabla}P}{\rho} \right.
\nonumber \\
    \left. +Z_2 \bm{\nabla}\frac{}{} \mu_1-Z_1 \bm{\nabla}\mu_2
   \right].
\label{e:d1a}
\end{eqnarray}

Quite generally, the chemical potential of ions $j$ is
$\mu_j=\mu_j^{(id)}+\mu_j^{(C)}$, where $(id)$ and $(C)$ label the
ideal gas and Coulomb contributions, respectively (see, e.g., Ref.\
\cite{hpy07}). Then $\bm{d}_1=\bm{d}_a+\bm{d}_b+\bm{d}_c$, with
\begin{eqnarray}
  \bm{d}_a&=&m_u Z_1 Z_2\,\frac{n_1 n_2}{n_e} \,
  \left( \frac{A_2}{Z_2}- \frac{A_1}{Z_1}
  \right)\,\frac{\bm{\nabla}P}{\rho},
\label{e:da} \\
  \bm{d}_b &=& \frac{n_1 n_2}{n_e} \,
  \left[ Z_2 \bm{\nabla}\mu^{(id)}_1-Z_1 \bm{\nabla}\mu^{(id)}_2\right]
\nonumber \\
  &=&\frac{k_BT}{n_e}\,\left(Z_2n_2 \bm{\nabla}n_1-Z_1n_1 \bm{\nabla}n_2 \right),
\label{e:db} \\
  \bm{d}_c &=& \frac{n_1 n_2}{n_e} \,\left[Z_2 \bm{\nabla}\mu^{(C)}_1-Z_1
  \bm{\nabla}\mu^{(C)}_2\right].
\label{e:dc}
\end{eqnarray}
In Eq.\ (\ref{e:db}) we have used the well-known relation
$\bm{\nabla}\mu^{(id)}_j=k_B T\,n_j^{-1}\bm{\nabla}{n_j}$.

Combined with (\ref{e:J1}), these equations give us the expression
for $\bm{J}_2$. It contains three terms labeled by subscripts $a$,
$b$ and $c$. The terms $a$ and $b$ are well known while the term $c$
seems new.

($a$). Assume that the matter is in hydrostatic equilibrium as a
whole. Then in Eq.\ (\ref{e:da}) we have
$\bm{\nabla}P=\rho\,\bm{g}$, so that $\bm{d}_a$ describes
gravitational sedimentation of the ions 2 (provided their effective
``molecular weight'' $A_2/Z_2$ is larger than that, $A_1/Z_1$, for
ions 1).

($b$). The term $\bm{d}_b$ is especially simple in the limit of $n_2
\ll n_1$. Then $n_e \approx Z_1 n_1$ and
$  \bm{d}_b =-k_BT \, \bm{\nabla}n_2 $
which corresponds to ordinary diffusion of ions 2. Generally, $\bm{d}_b$
describes diffusive motion of the ions if their number densities are
out of equilibrium.

($c$) The term $\bm{d}_c$ is most important in the regime of strong
ion coupling and can be accurately described in the ion-sphere approximation combined with
the linear mixing rule (e.g., Ref.\ \cite{hpy07} and references therein):
\begin{equation}
   \mu_j^{(C)}=-0.9 \,\frac{Z_j^{5/3}e^2}{a_e}, \quad
   \bm{\nabla} \mu_j^{(C)}=
   -0.3\,\frac{Z_j^{5/3}e^2}{a_e}\,\frac{\bm{\nabla}n_e}{n_e}.
\label{e:muC}
\end{equation}
Then
\begin{align}
  \bm{d}_c=0.3\,\frac{n_1 n_2}{n_e}\,\frac{Z_1 Z_2 e^2}{a_e} \,
  \left(Z_2^{2/3}-Z_1^{2/3} \right)\,\frac{\bm{\nabla}n_e}{n_e}.
\label{e:term3}
\end{align}
The structure of $\bm{d}_c$ is similar to that of $\bm{d}_a$: it
describes specific (``Coulomb'') sedimentation of ions 2 (provided $Z_2>Z_1$)
due to Coulomb coupling in the gravitational field.  Its remarkable feature is that
it operates even for ions with $A_1/Z_1=A_2/Z_2$. Such ions are
commonly thought to have the same ``molecular weights.'' Then
$\bm{d}_a=0$ (as long as we neglect small mass defects of ions 1 and
2) and one commonly assumes that such ions are not separated. We see
that it is not true.

By way of illustration let us rewrite Eq.\ (\ref{e:J1}) under
simplifying assumptions which are usually satisfied in the cores of
white dwarfs and outer envelopes of neutron stars. Assume that the
pressure is provided by strongly degenerate electrons, $P=P_e(n_e)$
[which allows us to express $\bm{\nabla}n_e$ in Eq.\ (\ref{e:term3})
through $\bm{\nabla}P$] and the hydrostatic equilibrium  is
established ($\bm{\nabla}P=\rho\bm{g}$). Then we obtain the
diffusion flux in the standard form

\begin{align}
   \bm{J}_2=D\,\frac{m_1 m_2 n}{\rho n_e}\,
   \left(Z_2n_2 \bm{\nabla}n_1-Z_1n_1 \bm{\nabla}n_2 \right)\nonumber \\
   + (\bm{u}_a+\bm{u}_c) m_2 n_2
\label{e:i2}
\end{align}
where

\begin{eqnarray}
  \bm{u}_a & = & \frac{\rho_1 n D}{\rho n_e k_B T}\,Z_1 Z_2 m_u
  \bm{g} \,\left( \frac{A_2}{Z_2}- \frac{A_1}{Z_1}
  \right),
\label{e:ua}
\\
  \bm{u}_c & = & \frac{\rho_1 n D}{ n_e k_B T}\,Z_1 Z_2
   \bm{g} \,\left(Z_2^{2/3}-Z_1^{2/3}\right)\,
  \frac{0.3 e^2}{a_eP\gamma}
\label{e:uc}
\end{eqnarray}
are the velocities of gravitational settling of ions 2 due to
``molecular weight'' difference and Coulomb separation, respectively;
$\gamma=\partial \ln P /\partial \ln \rho$.

Note that, when the matter is in hydrostatic equilibrium, the
gravitational settling of ions 2 is accompanied by ``lifting'' of
ions 1 (with $\bm{J}_1= -\bm{J}_2$). Such diffusive motion of ions
initiates collisional production of the specific entropy
($\dot{S}_{\mathrm{coll}}$) and the associated thermal energy release at a
rate $Q$ [erg~cm$^{-3}$ s$^{-1}$] (e.g., Refs.\  \cite{cc52,hirsh})
\begin{equation}
   Q=T\,\dot{S}_{\mathrm{coll}}=\frac{\rho}{\rho_1 \rho_2}\,\bm{J}_2 \bm{\cdot} \bm{d}_1
\label{e:Q}
\end{equation}
which is easily computed.

\section{Discussion and conclusions}

\begin{figure}[t]
\centering
\includegraphics[width=0.55\textwidth]{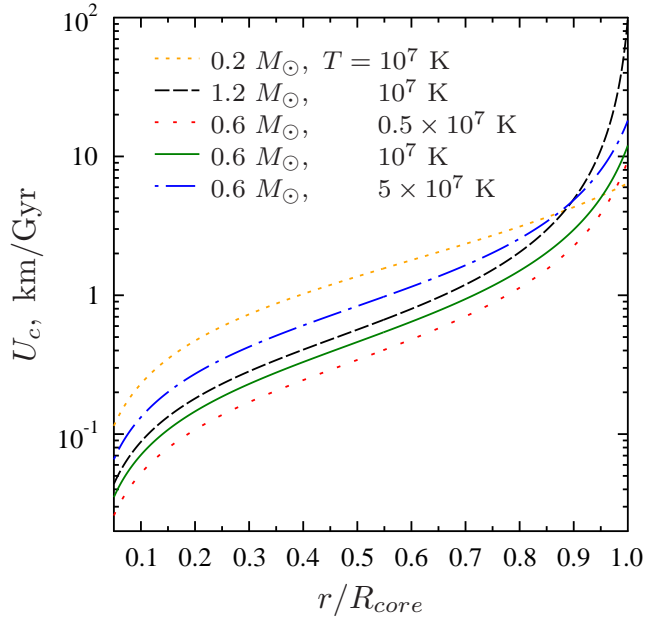}
\caption{(Color online) Velocity (\ref{e:uc}) of Coulomb settling in isothermal
degenerate cores ($r<R_{\mathrm{core}}$) of white dwarfs of different mass
(0.2, 0.6, and 1.2\,$M_\odot$, $M_\odot$ being the Sun's mass) and internal temperature
(0.5, 1, and 5 $\times 10^7$ K). We show
settling of $^{12}$C in mixture with $^{4}$He in $M=0.2\,M_\odot$ star. In other cases it is the settling
of $^{16}$O mixed with $^{12}$C. In all cases we assume that $n_1 = n_2$.}
\label{fig:wd}
\end{figure}

\begin{figure}[t]
\centering
\includegraphics[width=0.55\textwidth]{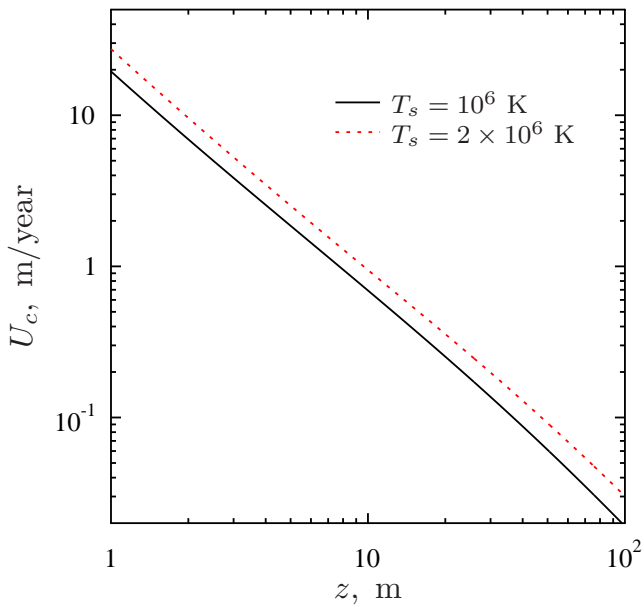}
\caption{(Color online) Settling velocity of $^{12}$C ions mixed with
$^4$He ($n_1=n_2$) in the outer envelope of a neutron star
(with gravity $g=2\times10^{14}$ cm~s$^{-2}$) versus depth $z$
(measured from the stellar surface)
for two effective surface temperatures $T_s=10^6$ and $2 \times 10^6$~K. While
$z$ varies from 1 to 100 m, the density increases from $\sim 10^5$ to
$\sim 10^9$ g~cm$^{-3}$.}
\label{fig:ns}
\end{figure}

Although the diffusion flux (\ref{e:i2}) has standard form, it
contains a new gravitational settling term (\ref{e:uc}) due to
Coulomb separation. This separation has been predicted by
Chang, Bildsten and Arras \cite{cb10} who considered equilibrium
distributions of ion mixtures including the Coulomb interaction term.
Thus we extend their work to nonequilibrium mixtures and show that
the Coulomb separation is pronounced in the
diffusion flux (\ref{e:i2}) and  drives
gravitational settling of ions.

The most pronounced effect occurs at
temperatures at which the ions constitute strongly coupled Coulomb
liquid. At lower temperatures the ions solidify and diffuse
much more slowly \cite{horowitz11}. At higher $T$ Coulomb coupling is weak and less
efficient (although generally available). The Coulomb sedimentation should be
especially important for the mixtures of ions with the same $A/Z$
(for instance, mixtures of $^4$He, $^{12}$C, and $^{16}$O ions). The
traditional gravitational sedimentation (\ref{e:ua}) in such mixtures
is greatly suppressed (can occur only due to mass defects of atomic
nuclei \cite{cb10}). The Coulomb settling in these mixtures (\ref{e:uc}) is
typically much stronger than (\ref{e:ua}). The ions with larger $Z$
should move to deeper layers. The effect is stronger for a larger
difference of $Z$ in the mixture.

Because the effect is primarily driven by gravitational forces, it
should be most pronounced in compact stars (white dwarfs, and
especially in neutron stars) with strongest gravity. First of
all we mean $^4$He--$^{12}$C cores of low mass white dwarfs and
$^{12}$C--$^{16}$O cores of more massive white dwarfs, and evolution
of similar mixtures in the envelopes of neutron stars. The velocity
of sedimentation is given by Eq.\ (\ref{e:uc})
using appropriate diffusion coefficients (e.g., Refs.\
\cite{hansenetal75,paquette,horowitz10,horowitz11}, and references
therein).

Figure \ref{fig:wd} presents the velocity $\bm{u}_c$ in the
$^{12}$C--$^{16}$O cores of medium mass and massive white dwarfs and
in the $^4$He--$^{12}$C cores of low mass white dwarfs. The adopted
temperate range $T \sim (0.5 - 5) \times 10^7$~K is appropriate to
rather old white dwarfs (e.g., Ref.\ \cite{hansen99}). The settling
velocities are higher for massive white dwarfs (with larger $g$). The
velocity profile throughout the core has a maximum at the core boundary
$r=R_{\mathrm{core}}$, where the gravitational acceleration $g(r)$  is the
largest. The velocity $\bm{u}_c\to 0$ as $r \to 0$ because $g(r) \to
0$ at the star's center. The maximum velocity in the massive
($1.2\,M_\odot$) white dwarf reaches $\sim 100$ km~Gyr$^{-1}$, meaning
that the $^{12}$C--$^{16}$O separation in the outer core can occur in
a few Gyrs. The velocity of Coulomb separation of $^{12}$C and
$^{16}$O ions is typically  lower than the settling velocity of
$^{22}$Ne ions in the  $^{12}$C--$^{16}$O core
\cite{isernetal91,bh01,db02,althausetal10,garciaetal10}, but the
fraction of   $^{22}$Ne ions is much smaller than the fractions of
$^{12}$C and $^{16}$O. Using Eqs. (\ref{e:uc}) and (\ref{e:Q}) we
have estimated the thermal energy generation rate $Q(r)$ which
accompanies this separation and found it insufficiently high to
noticeably reheat old white dwarfs. The profile $Q(r)$ has maximum in
the outer part of the white dwarf core. Note that our estimates neglect
the direct diffusion term [the first term in Eq.\ (\ref{e:i2})] which
can enhance $Q(r)$.

The Coulomb separation of  $^4$He, $^{12}$C, and $^{16}$O ions can be
important in isolated and accreting white dwarfs. It affects chemical
composition and, therefore, microphysics of white dwarf core (heat
capacity, thermal conductivity, neutrino emission, nuclear reaction
rates) as well as chemical, thermal, and nuclear evolution of white
dwarfs. Redistribution of ions due to Coulomb separation can
affect also vibration properties of stars (asteroseismology).

Coulomb separation of ions with equal $A/Z$ in neutron star envelopes
is much stronger than in white dwarfs. Figure \ref{fig:ns} plots the
sedimentation velocity $\bm{u}_c$ of $^{12}$C ions mixed with
$^{4}$He in the outer neutron star envelope versus depth $z$
(measured from the surface) for two effective surface temperatures,
$T_s$=1 and 2 MK. The temperature profile $T(z)$ within the envelope
has been determined by solving the heat transport equation for a
conserved heat flux emergent from stellar interior (see, e.g., Ref.\
\cite{cb10}). The envelope is nonisothermal and the temperature
gradient can affect diffusion which we ignore for simplicity.
Therefore, the presented curves should be treated as illustrative.
For the densities of $\sim 10^5 - 10^7$ g~cm$^{-3}$ (a few to a few
tens of meters under the surface) the sedimentation velocity can
reach a few meters per year. The separation can  affect nuclear
evolution of the matter in the outer layers of accreting neutron
stars.  It will change the thermal conductivity of this matter,
influence the relation between the surface and inner temperatures of
neutron stars and affect cooling of isolated and accreting neutron
stars (see, e.g., Refs.\ \cite{cb03,cb04,cb10, pcy97,py01,yp04}, and
references therein).

Similar Coulomb separation can occur in dusty plasmas which have many
applications in science and technology (e.g. Ref.\ \cite{dusty}).

\begin{acknowledgements}
We are grateful to A. I. Chugunov and A. Y. Potekhin for useful
discussions. D. G. Y. acknowledges support from RFBR (Grants No. 11-02-00253-a
and No. 13-02-12017-ofi-M), RF Presidential Program NSh 4035.2012.2, and
Ministry of Education and Science of Russian Federation (Agreement
No. 8409, 2012).
\end{acknowledgements}

\end{document}